\documentclass[prd,twocolumn,floats,floatfix,showpacs,nofootinbib]{revtex4}
\usepackage{graphicx}
\usepackage{dcolumn}
\usepackage{bm}
\usepackage{graphics}
\usepackage{slashed}
\usepackage{amssymb}
\usepackage{natbib}
\usepackage{amsmath}
\usepackage{url}
\usepackage{color}
\newcommand{\bea}{\begin{eqnarray}}
\newcommand{\ena}{\end{eqnarray}}
\newcommand{\bean}{\begin{eqnarray*}}
\newcommand{\enan}{\end{eqnarray*}}

\begin{document}

\title{A gravitational mechanism which gives mass to the vectorial
bosons}

\author{M. Novello\footnote{M. Novello is Cesare Lattes ICRANet Professor}} \email{novello@cbpf.br}

\affiliation{Instituto de Cosmologia Relatividade Astrofisica ICRA -
CBPF\\ Rua Dr. Xavier Sigaud, 150, CEP 22290-180, Rio de Janeiro,
Brazil}

\date{\today}

\begin{abstract}
A mechanism in which gravity is the responsible for the generation
of the mass of scalar bosons and fermions was proposed recently
\cite{novello}. The purpose of the present paper is to extend this
scheme to vector bosons. The main characteristic of this mechanism
is that the gravitational field acts really as a catalyzer, once the
final expression of the mass we obtained depends neither on the
intensity of the gravitational field nor on the value of Newton\rq s
constant.

\end{abstract}

\maketitle

\section{Introduction}

In the realm of high-energy physics, the Higgs model produced an
efficient scenario for generating mass to the vector bosons
\cite{halzen}. At its  origin appears a process relating the
transformation of a global symmetry into a local one and the
corresponding presence of vector gauge fields. The Higgs mechanism
(HM) appeals to the intervention of a scalar field $\varphi$ that
appears as the vehicle which provides mass to the  gauge vector
field  $ W_{\mu}.$ A particular form of self-interaction  of  $
\varphi$ allows the existence of a constant value $ V(\varphi_{0}) $
that is directly related to the mass of $ W_{\mu}.$ For the mass to
be a real and constant value (a different value for each field) the
scalar field  must be in a minimum state of its potential. This
fundamental state of the self-interacting scalar field has an energy
distribution described as
$$ T_{\mu\nu} = V(\varphi_{0}) \, g_{\mu\nu}$$
defining a cosmological constant of dimension $ length^{- 2} $ given
by $ \Lambda
 = \kappa\, V(\varphi_{0})$  that produces a particular configuration of
 the geometry of space-time through the equations of general
 relativity.

However, in the Higgs mechanism, this fact is not further analyzed,
once at that level, gravity is ignored. The reason for this is
attributed to the smallness of Newton\rq s constant $ \kappa = 8\pi
G_{N}/c^{4}.$ In the present paper, we shall argue in the opposite
way and will present a mechanism in which gravity is the true
responsible for the generation of the mass. The final expression of
the mass depends neither on the intensity of the gravitational field
nor on the value of Newton\rq s constant.

In our model we use a slight modification of Mach principle. We
start by recalling  Mach\rq s idea as the statement according to
which the inertial properties of a body $\mathbb{A }$ are determined
by the energy-momentum throughout all space. The simplest way to
implement this idea is to consider the state that takes into account
the whole contribution of the rest-of-the-universe onto $\mathbb{A
}$ as the most homogeneous one. Thus it is natural to relate it to
what Einstein attributed to the cosmological constant or, in modern
language, the vacuum of all remaining bodies. This means to describe
the energy-momentum distribution of all complementary bodies of
$\mathbb{A }$ in the Universe as
\begin{equation}
T^{U}_{\mu\nu} = \frac{\lambda}{\kappa} \, g_{\mu\nu}
\label{17abril}
\end{equation}
where $ \lambda $ is a constant of dimensionality $ (length)^{- 2}.$
In this work the body $\mathbb{A }$ will be identified with a vector
field interacting only gravitationally. Although we borrow the idea
of the rest-of-the-universe from Mach\rq s principle, and use its
correspondent form of energy-momentum distribution characterized by
a constant $ \lambda, $ as in (\ref{17abril}), one should not
identify naively this $ \lambda$ with the cosmological constant
obtained from actual observations on cosmology. In our scenario the
rest-of-the-universe should not be identified with an unique
structure, but instead it designates the environment of $ \mathbb{A}
$ or, in other words, all the remaining bodies in the universe that
can have an influence on $ \mathbb{A}. $ We synthesize the overall
effect of this structure on $ \mathbb{A} $ as an attribute of $
\lambda.$

\section{The vector field}
We start with a scenario in which there are only three ingredients:
a vector field, the gravitational field and an homogeneous
distribution of energy - that is identified with the vacuum. The
theory is specified by the Lagrangian (we use units $\hbar = c = 1$)

\begin{equation}
L = - \frac{1}{4} \, F_{\mu\nu} \, F^{\mu\nu} + \frac{1}{\kappa} \,
R - \frac{\lambda}{\kappa} \label{13abril1}
\end{equation}
The corresponding equations of motion are

$$F^{\mu\nu}{}{}_{; \nu} = 0 $$

and

$$ \alpha_{0} \, ( R_{\mu\nu} - \frac{1}{2} \, R \, g_{\mu\nu} ) = -
T_{\mu\nu} $$ where $F_{\mu\nu} = \nabla_{\nu} W_{\mu} -
\nabla_{\mu} W_{\nu}$ and, for graphical simplicity, we set  $
\alpha_{0} \equiv 2 / \kappa.$ In this theory, the vacuum $ \lambda
$ is invisible for $ W_{\mu}.$ The energy distribution represented
by $ \lambda$ interacts with the vector field only indirectly once
it modifies the geometry of space-time. In HM this vacuum is
associated to a fundamental state of a scalar field $ \varphi$ and
it is transformed in a mass term for $ W_{\mu}.$  The field
$\varphi$ has not only this intermediary role. Its function goes far
beyond this: it provides the vacuum energy. Indeed, in HM, one
identifies $ \lambda$ with the value of the potential $ V(\varphi)$
in its homogeneous state. To be able to realize this enterprize a
real scalar field must couple non-minimally with the vector field.

So much for a well-known scheme. Let us turn now to a new one. The
first step is to answer the following question: is it really
necessary to invoke a new field $ \varphi$ to act as a bridge
between the rest-of-the-universe vacuum and the vector field? We
claim that this is not necessary and we shall demonstrate this using
the universal character of gravitational interaction to generates
mass for $ W_{\mu} $ without introducing any extra field.

 The point of departure is the recognition that gravity may be
 the real responsible for breaking the gauge symmetry. For this, we modify the
 above Lagrangian to represent non-minimal coupling of the field $
 W_{\mu} $  and gravity in order to explicitly break such
 invariance. There are only two possible ways for this
 \cite{novellobscg} and the total Lagrangian must be of the form

\begin{eqnarray}
\mathbb{L} &=& - \frac{1}{4} \, F_{\mu\nu} \, F^{\mu\nu} + \frac{1}{\kappa} \, R \nonumber \\
 &+&  \frac{\sigma}{6} \,  R \, \Phi  + \sigma  \,
R_{\mu\nu} \, W^{\mu} \, W^{\nu} \nonumber \\
 &-&
\frac{\lambda}{\kappa} \label{210}
\end{eqnarray}
where we set

$$ \Phi \equiv  W_{\mu} \, W^{\mu}. $$

 The first two terms of  $ \mathbb{L} $
represents the free part of the vector and the gravitational fields.
The second line represents the non-minimal coupling interaction of
the vector field with gravity. The parameter $ \sigma $ is
dimensionless. The vacuum -- represented by $\lambda$ -- is added by
the reasons presented above and it must be understood as the
definition of the expression "the influence of the
rest-of-the-universe on $W_{\mu}".$ We will not make any further
hypothesis on this \cite{14abril}.

 In the present proposed mechanism, such $
\lambda $ is the real responsible to provide mass for the vector
field. This means that if we set $ \lambda = 0$ the mass of $
W_{\mu} $ will vanish.

Independent variation of $W_{\mu} $ and $g_{\mu\nu}$ yields
\begin{equation}
F^{\mu\nu}{}{}_{; \nu} + \frac{\sigma}{3} \, R  \, W^{\mu} + 2
\sigma \, R^{\mu\nu} \, W_{\nu} = 0 \label{8abril1}
\end{equation}

\begin{equation}
\alpha_{0} \, ( R_{\mu\nu} - \frac{1}{2} \, R \, g_{\mu\nu} ) = -
T_{\mu\nu}
 \label{8abril2}
\end{equation}
The energy-momentum tensor defined by
 $$T_{\mu\nu} = \frac{2}{\sqrt{- g}} \, \frac{\delta ( \sqrt{-g} \,
 L)}{\delta g^{\mu\nu}} $$
is given by
\begin{eqnarray}
T_{\mu\nu} &=& E_{\mu\nu} \nonumber \\
&+& \frac{\sigma}{3} \,  \nabla_{\mu} \nabla_{\nu} \Phi -
\frac{\sigma}{3} \, \Box  \Phi \, g_{\mu\nu} + \frac{\sigma}{3} \,
\Phi \, ( R_{\mu\nu} - \frac{1}{2} \, R \,
g_{\mu\nu} ) \nonumber \\
&+& \frac{\sigma}{3} \, R W_{\mu} \, W_{\nu}  + 2 \sigma
R_{\mu}{}^{\lambda} \, W_{\lambda} \, W_{\nu} + 2 \sigma
R_{\nu}{}^{\lambda} \, W_{\lambda} \, W_{\mu}  \nonumber
\\ &-& \sigma \, R_{\alpha\beta} \, W^{\alpha} \,W^{\beta} \,
g_{\mu\nu} - \sigma \, \nabla_{\alpha} \, \nabla_{\beta} \, (
W^{\alpha} \, W^{\beta} ) \, g_{\mu\nu} \nonumber \\
&+& \sigma \, \nabla_{\nu} \, \nabla_{\beta} ( W_{\mu} \, W^{\beta})
+ \sigma \, \nabla_{\mu} \, \nabla_{\beta} ( W_{\nu} \, W^{\beta})
\nonumber \\ &+& \sigma \, \Box ( W_{\mu} W_{\nu} ) +
\frac{1}{\kappa} \,\lambda \, g_{\mu\nu} \end{eqnarray}

where
$$ E_{\mu\nu} =  F_{\mu\alpha} \, F^{\alpha}{}_{\nu} + \frac{1}{4}
F_{\alpha\beta} \, F^{\alpha\beta} \, g_{\mu\nu}
$$

Taking the trace of equation (\ref{8abril2}) we obtain
\begin{equation}
R =  2 \, \lambda - \kappa \, \sigma \, \nabla_{\alpha} \,
\nabla_{\beta} \, ( W^{\alpha} \, W^{\beta} )
\end{equation}
Then, using this result back into equation (\ref{8abril1}) it
follows
\begin{eqnarray}
F^{\mu\nu}{}{}_{; \nu} &+&  \frac{2\, \sigma \, \lambda}{3} \,
W^{\mu}
 \nonumber \\
 &-& \frac{\kappa \, \sigma^{2}}{3} \,  \nabla_{\alpha} \,
\nabla_{\beta} \, ( W^{\alpha} \, W^{\beta}) \, W^{\mu} \nonumber \\
&+& 2 \, \sigma \, R^{\mu}{}_{\nu} \, W^{\nu} = 0
\end{eqnarray}

The non-minimal coupling with gravity yields an effective
self-interaction of the vector field and a term that represents its
direct interaction with the curvature of space-time. Besides, as a
result of this process the vector field acquires a mass $ \mu $ that
depends on the constant $\sigma $ and on the existence of $ \lambda$
given by
\begin{equation}
\mu^{2} =  \frac{2}{3} \, \sigma \, \lambda \label{219}
\end{equation}
Note that the Newton\rq s constant does not appear in our formula
for the mass. If $ \lambda $ vanishes then the mass of the field
vanishes. This is precisely what we envisaged to obtain: the net
effect of the non-minimal coupling of gravity with $ W^{\mu} $
corresponds to a specific self-interaction of the vector field. The
mass of the field appears only if we take into account the existence
of the rest-of-the-universe --- represented by $ \lambda$
--- in the state in which this environment is on the
corresponding vacuum. The values of different masses for different
fields are contemplated in the parameter $\sigma.$

\section{acknowledgements}
I would like to thank FINEP, CNPq and Faperj for financial support.
I would like also to thank J. M. Salim for many enthusiastic
conversations on the subject of this paper and A. Dolgov for a
critical comment in a previous paper concerning the value of $
\lambda.$


\begin{thebibliography}{50}
\bibitem{14abril} There is not any compelling reason to identify this constant with the actual
cosmological constant or the value of the critical density $ 10^{-
48} Gev^{4}$ provided by cosmology.
\end{thebibliography}
\end{document}